\documentclass[aps,pre, reprint,amsmath,amssymb, superscriptaddress]{revtex4-1}
\usepackage[pdftex]{hyperref}
\usepackage[usenames, dvipsnames]{color}
\usepackage{amsmath}
\usepackage{amssymb}
\usepackage{graphicx}
\usepackage[utf8]{inputenc} 
\usepackage[T1]{fontenc}
\usepackage{mathtools}

\begin{document}

\title{Bond and site color-avoiding percolation in scale free networks}

\author{Andrea Kadović}
\affiliation{Division of Theoretical Physics, Ruđer Bošković Institute, Zagreb, Croatia}

\author{Sebastian M. Krause}
\affiliation{Faculty of Physics, University of Duisburg-Essen, Duisburg, Germany}

\author{Guido Caldarelli}
\affiliation{IMT School for Advanced Studies Lucca, Piazza San Francesco 19, 55100, Lucca, Italy}
\affiliation{CRN Institute for Complex Systems, via dei Taurini 19, 00185, Rome, Italy}
\affiliation{London Institute for Mathematical Sciences, 35a South Street Mayfair, W1K 2XF, London, UK}
\affiliation{Linkalab, Complex Systems Computational Laboratory, Viale Elmas, 142 09122, Cagliari, Italy}

\author{Vinko Zlatić}
\affiliation{Division of Theoretical Physics, Ruđer Bošković Institute, Zagreb, Croatia}
\affiliation{CRN Institute for Complex Systems, via dei Taurini 19, 00185, Rome, Italy}


\begin{abstract}

Recently the problem of classes of vulnerable vertices (represented by colors) in complex networks has been discussed, where all vertices with the same vulnerability are prone to fail together. Utilizing redundant paths each avoiding one vulnerability (color), a robust color-avoiding connectivity is possible. However, many infrastructure networks show the problem of vulnerable classes of \textit{edges} instead of vertices. Here we formulate color-avoiding percolation for colored edges as well. Additionally, we allow for random failures of vertices or edges. The interplay of random failures and possible collective failures implies a rich phenomenology.  A new form of critical behavior is found for networks with a power law degree distribution independent of the number of the colors, but still dependent on existence of the colors and therefore different from standard percolation. Our percolation framework fills a gap between different multilayer network percolation scenarios.
\end{abstract}

\maketitle

\section{Introduction}

Studies of percolation on complex networks have a rich history~\cite{Newman_book, caldarelli2007scale}. Applications allow to asses the robustness of complex systems representable as networks. Most complex networks, as for example infrastructure networks, have to be connected in order to function properly. Varying types of connectivity can be needed for the proper functioning of a system. A prime example are interdependent networks which are formed of a number of different types of networks, with interdependency links between the vertices of different  types. Connections are dependent in the sense that if a vertex of type $A$ is disconnected, then the vertex of type $B$ also fails. This connectivity concept raises a plethora of surprising critical phenomena and is invaluable to explain real world problems like electric power shortages~\cite{buldyrev2010catastrophic}. Another example of the sensitivity to the definition of connectivity is the so called $k$-connected percolation~\cite{glauche2003continuum, newman2008bicomponents} in which vertices are connected only if there exist at least $k$ independent paths among them. Yet another example is color-avoiding percolation~\cite{krause2016hidden, krause2017color, shekhtman2018critical}. In these papers every vertex in a network is colored with one color out of a certain set of colors. A pair of  vertices is connected only if \emph{for each} color there exists a path in-between those two vertices which \emph{avoids} that color. 

From a theoretical point of view, percolation on networks~\cite{goltsev2006k, dorogovtsev2006k} exhibits a number of features which are rare or absent in more conventional models of percolation on lattices. Discontinuous phase transitions~\cite{igloi2002first}, phase transitions with Berezinskii--Kosterlitz--Thouless singularity~\cite{bauer2005phase}, magnetic field effects~\cite{shekhtman2018critical}, explosive percolation~\cite{achlioptas2009explosive} or inequality of site and bond percolation~\cite{radicchi2015breaking} represent just some of the critical phenomena that naturally emerge in percolation on complex networks. 

Recently some of the authors developed color-avoiding percolation (CAP)~\cite{krause2016hidden, krause2017color, shekhtman2018critical} as a tool to study the robustness of networks to simultaneous failures of certain classes of vertices. Classes are defined by their common vulnerabilities (for example the type of critical software they share), and a common vulnerability of a whole class is represented by its own color. The goal of CAP is to find a set of vertices which remain connected no matter which vulnerability is activated, i.e. which stay connected whichever the color is deleted color in the network. We define that two vertices are color-avoiding connected (CAC), if for \emph{every} color there exists a path between them which \emph{avoids} that color (vulnerability). Recent studies on multiplex networks  \cite{buldyrev2010catastrophic,cellai2013percolation,baxter2014weak,hackett2016bond, PhysRevE.94.012303} share some similarities with CAP, because layers can be thought of as the colors in CAP. If we thought of colors as layers, the method could as well be called layer-avoiding percolation. The main difference between CAP and other multiplex percolation scenarios is the choice of connectivity definition, where for every layer an avoiding path is needed. This is less strict than asking for connectivity in every single layer, and more strict than allowing a combination of all layers for connectivity. 

In this paper we extend CAP to networks with colored edges. This is important for a number of different network applications. An illustrative example are networks of public transportation, where stations are represented as vertices and direct connections as edges. Connections can be operated by bus, underground or other means of transportation. Coloring the direct connections according to the connection type, a transportation network with colored edges is constructed. The different connection types are vulnerable to different failures -- buses are delayed during rush hours, while technical problems can affect large parts of the underground system at the same time. In a robust public transportation system, every transportation type has to be avoidable, such that connections can be served by alternative means of transportation. 

We add another important aspect to the discussion of CAP -- random failures of vertices or edges. In the example of public transportation, there are often some edge failures due to reconstruction works, which are independent of correlated failures. If there are too many reconstruction works affecting the bus connections, the underground can become indispensable. 

We present numerical and analytical results for random network ensembles of Erd{\"o}s-R\'enyi type and with power law degree distribution \cite{clauset2009power}. Four different scenarios are considered: \emph{(i)} edge coloring with bond percolation (random failures of edges), \emph{(ii)} edge coloring with site percolation (random failures of vertices), \emph{(iii)} vertex coloring with bond percolation, and \emph{(iv)} vertex coloring with site percolation. On Erd{\"o}s-R\'enyi networks, all four scenarios show the same critical behavior, identical to \cite{krause2016hidden, krause2017color}. The critical behavior of CAP on networks with power law degree distribution was not jet understood. Here we find a new kind of CAP critical behavior which does not depend on how many colors are avoided, but still does depend on the existence of colors. 
Here we also find that the breaking of a bond-site percolation universality in networks with null percolation threshold~\cite{radicchi2015breaking} is also valid for CAP. Regarding different coloring cases, we find that the number of the vertices which are CAC does not depend if colors are treated as an edge or an vertex property.

In section II., we define an algorithm for finding the largest subset of vertices which are color-avoiding connected. After explaining assumptions of our model in section III., we develop a theory for calculating the relative size of the color-avoiding giant component and its criticality for networks with edge coloring together with bond percolation in section IV. After a short discussion of Erd{\"o}s-R\'enyi networks, we calculate the relative size of the color-avoiding giant component for networks with power law degree distribution. This is done step by step in an iterative procedure, with largely differing results for different values of the power law exponent describing the degree distribution. 
Further in section V. we deal with the site percolation and show the breaking of site-bond percolation universality for CAP, as well as the case of vertex coloring. The conclusion follows in section VI.

\section{Color-avoiding connectivity}

To quantify how connectivity of a network depends on the coloring of vertices or edges, we study the \textit{color-avoiding giant component (CAGC)}, as it was already defined in \cite{krause2016hidden,krause2017color,shekhtman2018critical}. We say a pair of vertices is color-avoiding connected, if for \textit{every} untrusted color there exists a path in-between which \textit{avoids} that color. The CAGC is defined as the largest set of vertices where every pair is color-avoiding connected. The number of nodes in the CAGC can scale with the network size, such that a macroscopic fraction of a large network is color-avoiding connected. This motivates the name color-avoiding \emph{giant} component. 
 
In the case of vertex coloring, a pair of vertices is color-avoiding connected regardless of the colors of this pair of vertices -- only the colors of the vertices on the path in between the pair of vertices are important~\cite{krause2016hidden}. This definition enables color-avoiding connectivity, even if all colors of vertices are untrusted.  
An alternative scenario was studied as well, where all the vertices colored with the color of the starting and ending vertex are trusted \cite{krause2017color}. 

\begin{figure}[t]
\includegraphics[scale=0.18]{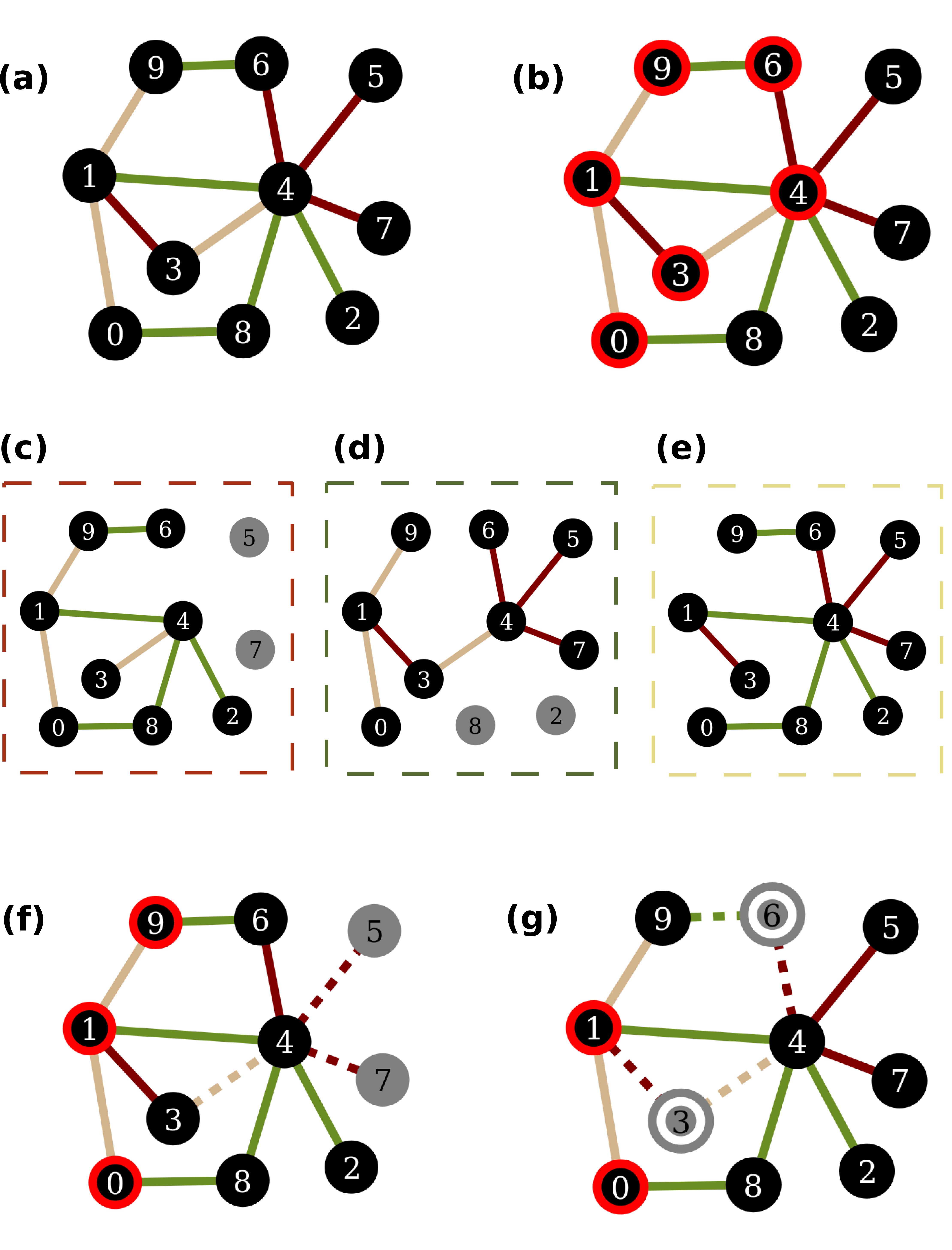}
\centering
\caption{\textbf{The color-avoiding component under the bond percolation and the site percolation.} Here we present a network with its edges colored in three different colors -- \textit{brown, green, yellow} \textbf{(a)}. The giant color-avoiding component associated with this network is designated with red rounded vertices in \textbf{(b)}. One can find the color-avoiding component as an intersection of connected components when avoiding each color individually. Color of dashed box represents the color which is avoided, for each color in network in \textbf{(c)-(e)}. Vertices which are not a part of connected component are colored in gray. Thus the black vertices in, for example \textbf{(c)}, are connected through the paths which are free of \textit{brown} edges. Note that vertices with only one edge, i.e. vertices with number $2,5,7$, do not belong in the color-avoiding component since they surely become isolated when color of their single edge is avoided. Therefore a vertex which belongs in giant color-avoiding component has to have degree equal or larger then 2. Comparing vertices number $6$ and $8$, one can notice that a vertex in color-avoiding component (vertex 6) has to have at least two edges of different colors. Deleted edges in bond percolation \textbf{(f)} and deleted vertices (vertices 6 and 3) with their edges in site percolation \textbf{(g)} are marked as dashed.}
\label{fig:the_problem}
\end{figure}

In networks with colored edges, such a distinction between different scenarios is not necessary. The CAGC in the presence of colored edges consists of all vertices which stay color-avoiding connected, and because vertices are not colored, they are always trusted (see fig.~\ref{fig:the_problem}(a-e)). 

It is important to stress that fig.~\ref{fig:the_problem} also describes the algorithm which we use to numerically study the problem of CAP. In other words we \emph{(i)} delete every color, \emph{(ii)} find the largest component surviving without the deleted color and \emph{(iii)} then find an intersection of all the largest components formed for each of the deleted colors. This algorithm works perfectly for colored edges, while for colored vertices we have to extend the algorithm as described in \cite{krause2016hidden}. 

Before, color-avoiding percolation was only accounting for correlated failures of all vertices with the same color~\cite{krause2017color}. Here we also allow for additional random failures of vertices or edges regardless of the color distribution. We study how standard percolation affects the size of the CAGC and present a condition for its existence in a given network. This allows us to study the critical behavior also on colored scale-free networks, which is remarkably different from colored Erd{\"o}s-R\'enyi networks~\cite{krause2017color}. It also expands our possibilities to dilute networks by choosing either vertices (site percolation) or edges (bond percolation) and to test if there is any difference between these two kinds of percolation (see fig.~\ref{fig:the_problem}(f-g)).

\section{Model}
        
We consider the configuration network model, where the degree $k$ of vertices is characterized with a degree distribution $p_k$ \cite{molloy1995critical, Newman_book, caldarelli2007scale}. The degree $k$ is the number of first neighbors of a given vertex. 
For defining the coloring of the network, we consider two different scenarios. \emph{(i)} A network with colored edges is constructed by assigning to every edge $e$ exactly one color $c_e\in \{ 1,2,\dots,C \}$.  The colors are chosen independently random, every color $c$ with a probability $r_c$. The probabilities of all colors sum to one, $\sum_{c=1}^C r_c=1$. For large networks, the share of edges with color $c$ converges to $r_c$. 
\emph{(ii)} A network with colored vertices is constructed by assigning to every vertex $v$ exactly one color $c_v\in \{ 1,2,\dots,C \}$ with probability $r_c$.

As we want to account for additional random failures of edges (or vertices), we use a dilution approach in which a fraction of edges (or vertices) $\phi$ is retained after uniform dilution of the network. Consequently $1-\phi$ is the fraction of edges (or vertices) which is removed from the network. This procedure also helps to better quantify the percolation threshold~\cite{Newman_book, radicchi2015predicting} on scale-free networks. Let $\mathcal{G}$ be the set of vertices in the largest component of the diluted network. In the thermodynamic limit of networks with infinite size, there is a percolation threshold $\phi_{crit}$ which separates a non-percolating phase for $\phi<\phi_{crit}$ from a percolating phase for $\phi>\phi_{crit}$~\cite{Newman_book}. $\mathcal{G}$ is finite in the non-percolating phase, while it includes a macroscopic fraction of vertices in the percolating phase. 

We further treat the additional effect of the edge coloring, where the first $T\leq C$ colors are untrusted and thus need to be avoidable. We define $\mathcal{T}=\{1,2,\dots T\}$ as the set of untrusted (avoidable) colors. Our theory will work even in the cases in which we choose not to avoid all the colors but only some subset of them. This is important because in some applications we could, for example, have a completely safe category which could be colored with the color which is therefore never avoided. Also the theory which includes avoidance of only a subset of colors is more general. Let $\mathcal{G}_{\rm color}$ be the set of vertices belonging to the CAGC. Similar to the case of colored vertices~\cite{krause2016hidden,krause2017color,shekhtman2018critical}, we construct $\mathcal{G}_{\rm color}$ but now starting from the $\phi$-diluted network. For every color $c$, we destroy all edges with color $t \in \mathcal{T}$ and find the set of nodes in the remaining largest component, denoted as $\mathcal{G}_{\{t\}}$. In the subscript, the $\{t\}$ denotes the avoided color. Every vertex pair in $\mathcal{G}_{\{t\}}$ is connected by a path without \emph{any} edge having color $t$. We finally find $\mathcal{G}_{\rm color}=\cap_{t=1}^T \mathcal{G}_{\{t\}}$.

If we have a vertex coloring instead of the edge coloring, we assume that all sending/receiving vertices are trusted, and only vertices on the path between a pair of nodes are avoided. Again, we start with the $\phi$-diluted network. For every color $c$, we destroy all vertices with color $c$ and find the set of nodes in the remaining largest component, denoted as $\mathcal{G}_{\{t\}}$. By adding all direct neighbors of vertices in $\mathcal{G}_{\{t\}}$, we obtain a larger set of vertices $\mathcal{G}_{\{t\}}^+$. Every vertex pair in $\mathcal{G}_{\{t\}}^+$ is connected by a path without \emph{any} vertex \emph{in between} having color $c$. We finally find $\mathcal{G}_{\rm color}=\cap_{t=1}^T \mathcal{G}_{\{t\}}^+$.

\subsection{Color-avoiding critical threshold}

The color-avoiding percolation threshold $\tilde{\phi}_{crit}$ characterizes the formation of the CAGC. It depends on $\phi_{crit}$, because every giant color-avoiding component must be a subset of the regular giant component. 
As the CAGC is the intersection of largest components $\mathcal{G}_{\{t\}}$ (respectively $\mathcal{G}_{\{t\}}^+$ for the case of colored vertices) for different avoided colors $t \in \mathcal{T}$, let us first discuss the critical onset for $\mathcal{G}_{\{t\}}$ (respectively $\mathcal{G}_{\{t\}}^+$), which is identical to the color-avoiding percolation threshold for a single avoided color. 
It can be modeled as a \emph{thinning stochastic process}~\cite{durrett2010probability, stumpf2005subnets, arratia2014scale} of an arbitrary initial degree probability $p_k$ to a thinned one $\tilde{p}_k$, i.e. color-avoiding one, when color $t \in \mathcal{T}$ is thinned out or avoided,
\begin{equation}
    \tilde{p}_k = \sum\limits_{l = k}^{\infty} p_l \binom{l}{k} (1-r_t)^k r_t^{l-k}.
\end{equation}
The generating functions associated with initial degree distribution are defined as $g_0(x) = \sum_{k = 0}^{\infty} p_k x^k$, and for the color-avoiding degree distribution we use $\tilde{g}_0 (x)= \sum_{k = 0}^{\infty} \tilde{p}_k x^k$  in analogy with $ \tilde{p}_k$. The thinning of degree distribution in a sense of avoiding color means that
\begin{equation}
    \tilde{g}_0 (x) = g_0 \left( (1-r_t) x + r_t  \right).
\end{equation}
Since the average degree and other higher moments, noted with $\left\langle k^m \right\rangle = \sum_{k = 0}^{\infty} k^m p_k, \ m \in \mathbb{N}$ and $\left\langle \tilde{k}^m \right\rangle$ associated with $\tilde{p}_k$, are given as a derivations of a generating function~\cite{Newman_book}, it is easy to show that 
\begin{align}
    \tilde{\left\langle k \right\rangle} &= \left\langle k \right\rangle (1-r_t), \\
    \left\langle \tilde{k}^2 \right\rangle - \left\langle \tilde{k} \right\rangle &=  \left[ \left\langle k^2 \right\rangle -\left\langle k \right\rangle \right] (1-r_t)^2.
\end{align}
This is specifically needed to compute the percolation thresholds of the Erd\"os-R\'enyi and scale free networks. We remind that the Erd\"os-R\'enyi network has a Poisson degree distribution $p_k = {\left\langle k \right\rangle}^k \mathrm{e}^{-\left\langle k \right\rangle} / k!, \ k \geq 0$ and a percolation threshold at $\phi_{crit} = 1 / \left\langle k \right\rangle$~\cite{Newman_book}. The scale free network is defined with power law degree distribution $p_k = A k^{-\gamma}, k \geq 1$, and has percolation threshold at $\phi_{crit} = \left\langle k \right\rangle / \left\langle k(k-1) \right\rangle$~\cite{Newman_book}. Since the avoiding of color repeats for every avoidable color, (for example see fig.~\ref{fig:the_problem}(c-e)), the assumption is that the smallest of all color-avoiding components $\mathcal{G}_{\{t\}}$ (or $\mathcal{G}_{\{t\}}^+$ in the case of vertex coloring) will be avoiding the dominant color $t$ with largest $r_t$. Consequently, this component will have the largest percolation threshold for creation of the giant connected component. Thus the percolation threshold for the formation of CAGC is a function of the probability $1-\mathrm{max}_{t \in \mathcal{T}} r_t$ of avoiding the dominant color~\cite{krause2017color}. 
Finally,
\begin{equation}
    \tilde{\phi}_{crit} = \phi_{crit} \ \frac{1}{1- \mathrm{max}_{t \in \mathcal{T}} r_t},
\label{eq:threshold}
\end{equation}
with which we confirm the percolation threshold for the Erd\"os-R\'enyi network and expand it to the  mean field regime of the scale-free network with pure power law degree distribution.

\section{Color-avoiding bond percolation}
    
Our calculation follows the essence of the regular (bond) percolation on noncolored networks~\cite{newman2001random,burda2001statistical,cohen2001breakdown,cohen2002percolation,dorogovtsev2008critical}. We adapt these approaches to the color-avoiding percolation, by including the possibility that all edges of particular color fail at once~\cite{krause2017color}.  Site percolation will be studied in a later section. As in previously studied percolation models, we use generating functions for the degree distribution of the initial network $g_0(x)=\sum_{k = 0}^{\infty} p_k x^k$, as well as for the excess degree distribution~\cite{Newman_book} $g_1(x) = \left( \left\langle k \right\rangle \right)^{-1} \sum_{k = 1}^{\infty} k p_k x^{k-1}$. To calculate the size of the standard giant component, it is common to solve a self-consistent relation for the probability $u$ that an edge \textit{is not connecting} to it~\cite{Newman_book}. For color-avoiding percolation, we need to calculate a set of edge failure probabilities $u_{\mathcal{Q}}$, for a set of untrusted colors $\mathcal{Q} \subseteq \mathcal{T}$. The quantity $u_{\mathcal{Q}}$ defines the probability that an edge \textit{fails connecting to} the largest component $\mathcal{G}_{Q}$, obtained when all untrusted colors $q \in \mathcal{Q}$ are deleted in the same time. We denote the relative size of the CAGC, avoiding the colors in the set $\mathcal{T}$, as $B_{color}^{\mathcal{T}}$. It can be computed in a lengthy calculation using generating functions together with event negations and the inclusion-exclusion principle~\cite{krause2017color}: 
\begin{align}
    B_{color}^{\mathcal{T}} = 1+ \sum\limits_{\mathcal{Q} \subseteq \mathcal{T}} &\left( -1 \right)^{\mid \mathcal{Q} \mid} g_0 (u_{\mathcal{Q}}). 
\label{eq:B_color}
\end{align}

We exclude the empty set from $\mathcal{T}$ in eq.~\ref{eq:B_color}, since its solution is trivial ($u_{\varnothing} = 1$). In this approach, instead of only one variable $u$ needed to compute the giant connected component of ordinary percolation, $2^{|\mathcal{T}|}-1$ variables $u_{\mathcal{Q}}$ are needed, where $\mathcal{Q}$ takes on all non-empty subsets of $\mathcal{T}$. 
 
The inherent self-consistent equations for the variables $u_{\mathcal{Q}}$ are
\begin{align}
    u_{\mathcal{Q}} = (1-\phi) + &\phi \left[ \sum\limits_{q \in \mathcal{Q}} r_q g_1(u_{\mathcal{Q} \setminus \{ q \} }) \right. \nonumber \\
    &\left. + \left(1- \sum\limits_{q \in \mathcal{Q}} r_q\right) g_1(u_{\mathcal{Q}}) \right]. 
\label{eq:self_consistent}
\end{align}
These equations are valid either for colored edges (with edge color shares $r_q$) or for colored vertices (with vertex color shares $r_q$). 
The brackets $\{\}$ represent an avoiding set of specified color designated with a small letter, while large letters are used for the names of sets of avoidable colors. The first term simply describes the probability $1-\phi$ that a given edge is removed. The prefactor $\phi$ of the second term describes the probability that the edge is not removed. The first term in brackets $[...]$ is a conditional probability: Given that the edge still exists, we sum over all the colors $q$ in the set of avoided colors $Q$ the probability that the edge (or the reached vertex) is of color $q$, which therefore cannot be avoided. Additionally, \emph{none} of the outgoing edges from the reached vertex is connecting to \emph{any} of the components $\mathcal{G}_{\{c\}}$ avoiding colors $c \in \mathcal{Q} \setminus \{q\}$. The second term in the brackets $[...]$ covers another conditional probability: Given that the edge still exists, we calculate the probability that the edge (or the reached vertex) has a color different from all avoided colors in $\mathcal{Q}$ and that at the same time \emph{none} of the outgoing edges from the reached vertex is connecting to \emph{any} of the components $\mathcal{G}_{\{q\}}$ avoiding the colors $q \in\mathcal{Q}$. In this way, a condition is fulfilled that the edge connects to \emph{none} of the components $\mathcal{G}_{\{q\}}$ avoiding colors $q \in \mathcal{Q}$. Consequently, the self-consistent equations must be solved iteratively from avoiding single colors to avoiding the whole set $\mathcal{T}$ of avoidable colors.

\subsection{Critical exponents}

Near the percolation threshold $\tilde{\phi}_{crit}$, we expect a power-law behavior~\cite{cardy1996scaling,krause2016hidden} of the form 
\begin{align} 
\label{eq:critical_onset}
B_{color}^{ \mathcal{T}} \sim \left( \phi - \tilde{\phi}_{crit} \right)^{\beta_B^{ \mathcal{T}}}.
\end{align}

The critical exponent can be found with an expansion according to small quantities~\cite{cohen2002percolation}. We see that the critical behavior implies $B_{color}^{ \mathcal{T}}\to 0$ with $\phi\to\tilde{\phi}_{crit}$. Accordingly, we expect that the single edge failure probability is close to one at the critical point, $u_{\mathcal{Q}}\to 1$. Here we will use the dual picture~\cite{krause2017color} with probabilities $v_{\mathcal{Q}}$ that a link \emph{connects to all} $\mathcal{G}_{\{q\}}$. Using the dual variables $v_{\mathcal{Q}}$  allows us to rephrase the problem, entirely based on small quantities: 
\begin{align} 
\label{eq:vQ_2_uP}
    v_{\mathcal{Q}} &= 1 + \sum\limits_{\mathcal{P} \subseteq \mathcal{Q}} \left(-1 \right)^{\mid \mathcal{P} \mid} u_{\mathcal{P}},\\ 
\label{eq:uQ_2_vP}
    u_{\mathcal{Q}} &= 1 + \sum\limits_{\mathcal{P} \subseteq \mathcal{Q}} \left(-1 \right)^{\mid \mathcal{P} \mid} v_{\mathcal{P}}\equiv 1-\epsilon_{\mathcal{Q}},\\ 
\phi &= \tilde{\phi}_{crit} + \delta , \qquad v_{\mathcal{P}}, \epsilon_{\mathcal{Q}},\delta \ll 1. \label{eq:phi_c_delta}
\end{align}
One important special feature of $v_{\mathcal{Q}}$ arises if all colors are avoided: If $\mathcal{Q}=\{1,2,\dots,C\}$ then $v_{\mathcal{Q}}=0$ \cite{krause2017color}, because a single edge cannot avoid all colors. 

The dual picture with small variables allows us to expand $g_0(u_{\mathcal{Q}})$ in eq.~\ref{eq:B_color} around $g_0(1)$. For Poisson graphs, we use $g_0(x)=g_1(x)=\exp(\left\langle k \right\rangle (x-1))$ and find 
\begin{align} 
\label{eq:poi_gen_tay}
    g_0(1-\epsilon)&=\exp(- \left\langle k \right\rangle \epsilon) \nonumber\ \\
    &\approx 1- \left\langle k \right\rangle \epsilon + \frac{\left\langle k \right\rangle ^2}{2}\epsilon^2\,. 
\end{align}
Plugging into eq.~\ref{eq:B_color} and using $r_1=r_2=\dots=r_{|\mathcal{T}|}$ as well as $v_{\{1\}}=v_{\{2\}}$ etc., many terms cancel out and we find 
\begin{align} 
    B_{color}^{\mathcal{T}} &= \left\langle k \right\rangle v_{\mathcal{T}} \nonumber\\
    &\quad + \frac{\left\langle k \right\rangle ^2}{2}\sum\limits_{t=1}^{|\mathcal{T}|-1} \binom{|\mathcal{T}|}{t} v_{\{1,\dots,t\}}v_{\{t+1,\dots,\mathcal{T}\}} \nonumber\\
    &\quad +\dots,
\label{eq:poi_expansion}
\end{align}
where $|\mathcal{T}|$ is the number of avoided colors (non-empty elements) in set $\mathcal{T}$. Notice that the linear expansion in $\epsilon$ of $g_0(1-\epsilon)$ is not enough if all colors are avoided, because then $v_{\mathcal{T}}=0$. Therefore, we need higher order terms to describe the critical behavior, which is typically not necessary for other percolation types~\cite{cohen2002percolation}. 

In the following, we discuss the scaling of $v_{\mathcal{Q}}$ with $\delta$ to complete the discussion of critical exponents for Poisson graphs. For scale free graphs, higher derivatives of the generating function diverge, what makes it impossible to use the Taylor expansion. This problem can be overcome based on an asymptotic expansion.

\subsection{Poisson graphs}

\begin{figure}
\includegraphics[trim=45 0 20 0,clip,scale=0.36]{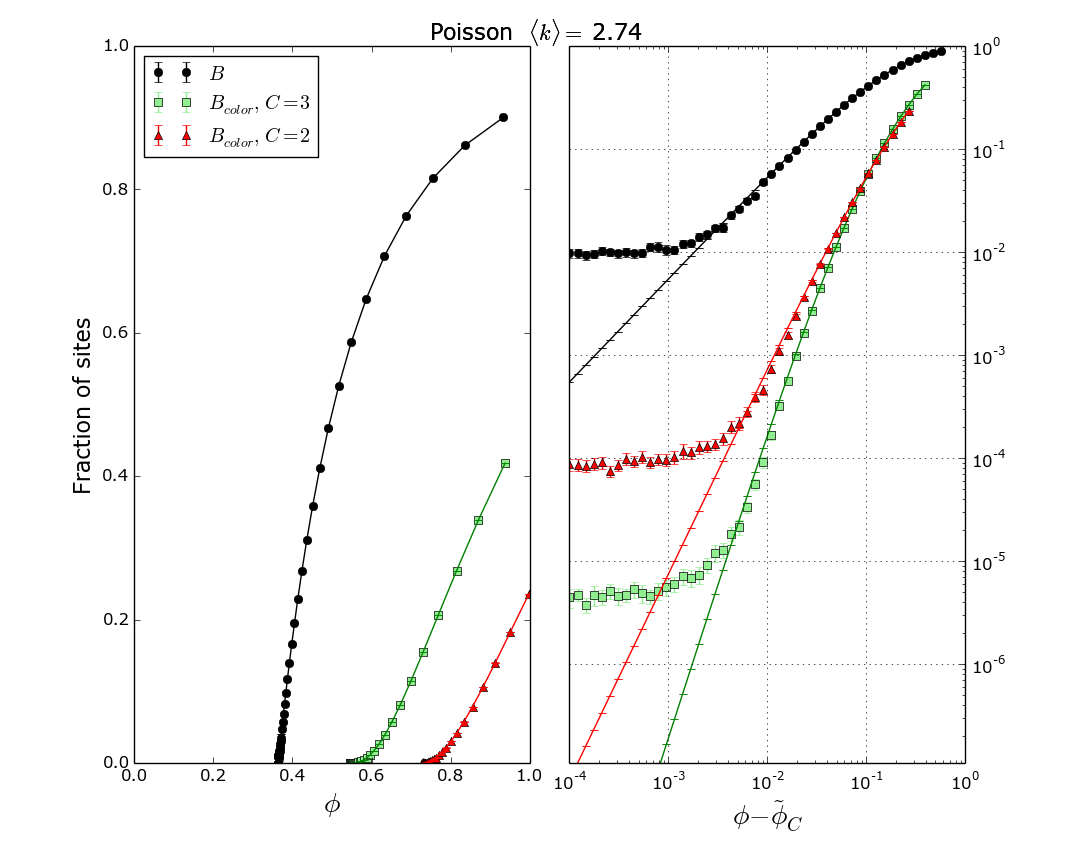}
\centering
\caption{\textbf{Erd\"os-R\'enyi edge colored network with equal color frequencies under bond percolation when all colors are avoided.} The relative size of the giant connected component is marked with \textit{black dots} and for the color-avoiding connected component CAGC two cases are shown -- \textit{red triangles} for $C= 2$ and \textit{green squares} for $C= 3$ avoiding colors. Except for the results from simulations which are shown as symbols, the results based on our theory are shown in \textit{lines} of the associated color. Numerical averages are done over $150$ networks of size $N=10^6$. Analytical result are obtained through $200$ iterations of self consistent equations presented in this paper. Standard deviations are shown as barely visible error-bars for both analytical and numerical results. For both cases, edges are colored with identical shares of the colors ($r_1=\dots=r_C$), all colors are avoided. Parameter $\phi$ is relative number of edges which stay after uniform bond dilution, starting from network with Poisson degree distribution with average degree $\langle k \rangle = 2.74$. Our theory corresponds with the numerical results which are strongly color dependent for all values of the parameter $\phi$. On the right, results are repeated with logarithmic scaling. The critical exponent is equal to the number of avoided colors.}
\label{fig:poisson}
\end{figure}

The relative size of the CAGC is completely defined by solving the set of self-consistent equations \ref{eq:self_consistent} and \ref{eq:B_color}. Our analytical results for networks with Poisson degree distribution are shown in fig.~\ref{fig:poisson} (shown with the red line for two avoided colors, and the green line for three avoided colors), accompanied with numerical results (symbols with error bars of one standard deviation). The algorithm for the numerical estimation of the relative size of the CAGC is sketched in fig.~\ref{fig:the_problem}: For every avoided color $t \in \mathcal{T}$, the largest remaining component $\mathcal{G}_{\{t\}}$ is identified and finally the intersection over all these components is used to identify vertices in $\mathcal{G}_{\rm color}$. We use an average over $150$ network configurations of the canonical ensemble (fixed number of vertices $N = 10^6$ and degree distribution $p_k$), with colors on the edges as an annealed degree of freedom and identical color probabilities $r_1=r_2=\dots=r_C$. The colors are newly assigned in every configuration (uniformly random distributed), after the vertices are rewired. Every color is treated as avoidable ($\mathcal{T} = \{1,...,C\}$). Simulations are done with the Graph-tool library in Python~\cite{peixoto2014graph}. We checked that the plateau for small $\phi$ in the numerical results on the right sub-figure is consequence of the finite size of simulated networks. The color-avoiding component is more rigid then the standard giant component (shown with the black line and symbols) and the CAGC is greatly affected by the number of colors. We emphasize that the CAGC has exactly the same size if vertices are colored instead of the edges, because both versions result in the same equations \ref{eq:self_consistent} and \ref{eq:B_color}, just with altered meaning of the color shares $r_q$. This result is tested with numerical results, which are not shown for Poisson graphs, but for scale-free graphs we will discuss it below in fig.~\ref{fig:v_e_color}. Differences in the critical behavior for different numbers of avoided colors are evident on the right of fig.~\ref{fig:poisson}, where the results are repeated in double-logarithmic scaling. The critical exponent is $\beta_{B}^{\mathcal{T}}=C$. This is identical with the result reported before~\cite{krause2016hidden, krause2017color}, however here we confirm this result for colored edges.

For the scaling of $v_{\mathcal{Q}}$, we find with identical color shares of all colors
\begin{align}
v_{\mathcal{Q}} &\sim \delta^{|\mathcal{Q}|}, 
\end{align}

where $|\mathcal{Q}|$ is number of avoided colors (elements) in set $\mathcal{Q}$. This result is found iteratively by plugging the Taylor expansion of $g_1(1-\epsilon)$ into eq.~\ref{eq:self_consistent}. Exactly the same procedure is described in detail in~\cite{krause2017color}. Plugging into eq.~\ref{eq:poi_expansion}, we are able to confirm the finding $\beta_{B}^{\mathcal{T}}=C$ (compare eq.~\ref{eq:critical_onset}).

\subsection{Scale free graphs}

\begin{figure}
\includegraphics[trim=40 0 30 0,clip,scale=0.4]{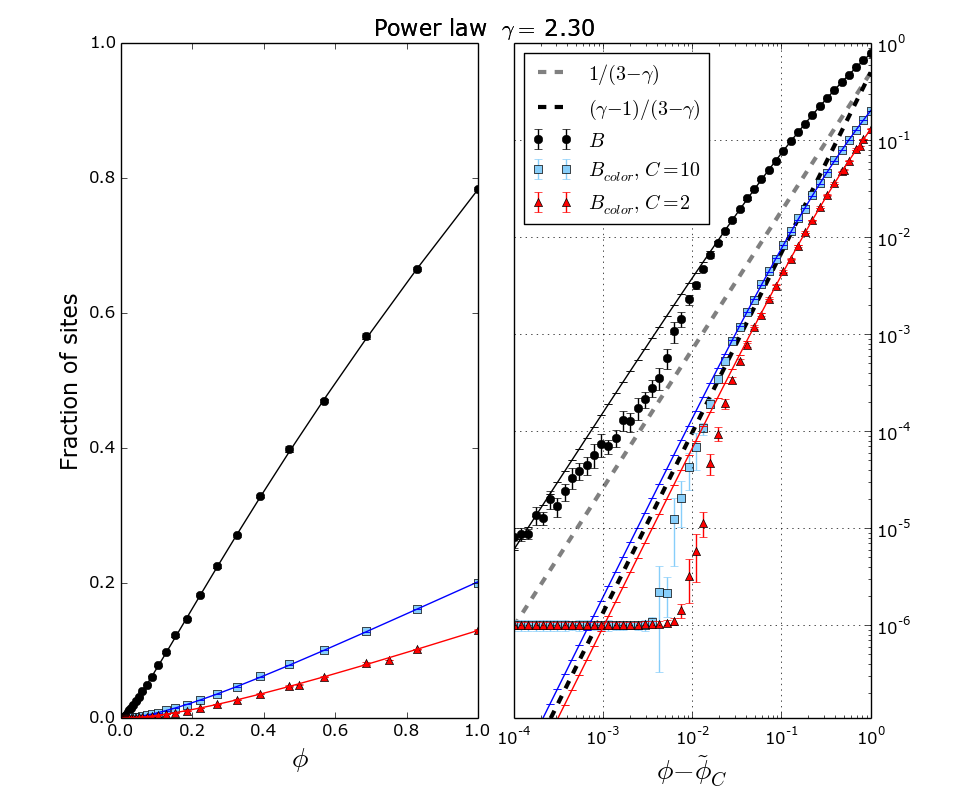}
\centering
\caption{\textbf{Scale-free edge colored network with equal color frequencies under bond percolation when all colors are avoided has color independent critical behavior.} Supplementing fig.~\ref{fig:poisson}, here we present results for a power law degree distribution with exponent $\gamma = 2.3$, with identical average degree as for the Poisson graph. \textit{Red triangles} show numerical results for the relative size of the CAGC with $C= 2$ avoided colors, \textit{blue squares} with $C= 10$ avoided colors. \textit{Black circles} show the relative size of the standard giant component. The corresponding \textit{solid lines} show analytical results with good agreement. On the right we see in logarithmic scaling that the critical exponent is independent of the number of avoided colors. \textit{The dashed black line} is a power law with exponent $(\gamma-1)/(3-\gamma)$ which is the critical exponent of CAGC that we have computed. \textit{The dashed gray line} is a power law with  exponent $1/(3-\gamma)$, which is a critical exponent of standard percolation on scale free networks. 
}
\label{fig:powerlaw}
\end{figure}

\begin{figure}
\includegraphics[trim=40 0 30 0,clip,scale=0.4]{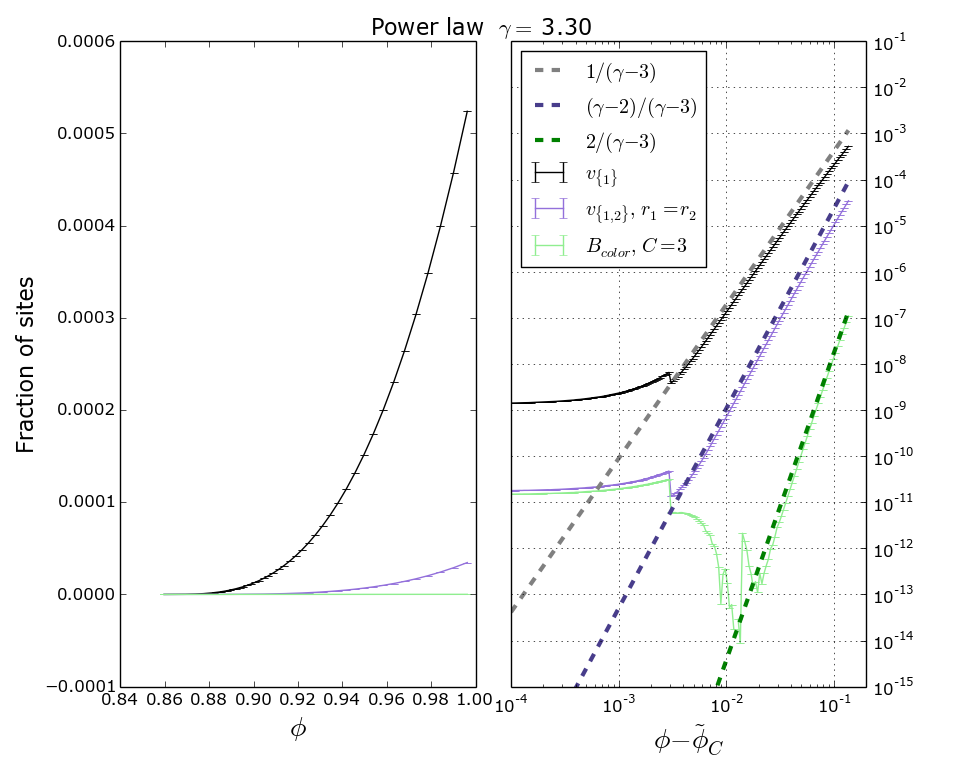}
\centering
\caption{
\textbf{Rich critical behaviour of color-avoiding percolation on scale-free edge colored network with equal color frequencies under bond percolation when all colors are avoided.} Here we present only analytic results for a power law degree distribution with exponent $\gamma = 3.3$ and $C= 3$. Characteristics of analytical calculation are the same as described in caption of fig.~\ref{fig:poisson}. \textit{Full lines} show results for iteratively solved equations, while \textit{dashed lines} on the right represent auxiliary lines for the critical exponents of CAGC (noted in \textit{green}) as well as for probabilities $v_{\{1\}}$ (in \textit{black}) and $v_{\{1,2\}}$ (in \textit{purple}). We see that the probability of connecting to the giant component which is avoiding two colors in same time is smaller then the probability of connecting to the giant component which avoids only one color, while they are both still behaving critically. However, the former one has the critical exponent $(\gamma-2)/(\gamma-3))$, which is larger then for the latter $1/(\gamma-3)$.  We also find a new critical exponent $2/(\gamma-3))$ for CAGC in this regime of power law exponent, but when we do not trust any color.
}
\label{fig:powerlaw_g33}
\end{figure}

Finding results for scale free graphs is more complicated than for Poisson graphs. Analytical results (lines) and numerical results (symbols) for networks with power law degree distribution with exponent $\gamma=2.3$ are shown in fig.~\ref{fig:powerlaw}. As for Poisson graphs, analytical results are computed with equations \ref{eq:self_consistent} and \ref{eq:B_color}, and numerical results are averages over $150$ network configurations of the canonical ensemble (fixed number of vertices $N = 10^6$ and power law degree distribution $p_k$). Every color is treated as avoidable ($\mathcal{T} = \{1,...,C\}$), all color shares are equal ($r_1=r_2=\dots=r_C$). We see that the CAGC for two avoided colors (red triangles with error bars of one standard deviation, and red line) and for ten avoided colors (blue squares and blue line) share the critical value with standard percolation (black circles and line) $\phi_{crit}=\tilde{\phi}_{crit}=0$. On the right of the figure, results are repeated in logarithmic scaling. Here we see that the critical exponent $\beta_{B}^{\mathcal{T}}=(\gamma-1)/(3-\gamma)$ is independent of the number of avoided colors. This surprising results are in sharp contrast to all results with Poisson graphs~\cite{krause2016hidden,krause2017color,shekhtman2018critical}, which were strongly depended on number of avoided colors.

The novelty of CAP which makes the criticality independent of the number of avoided colors, but still dependent on the existence of color on network also remains in scale free networks in regimes of power law exponents that are not mean field in nature. For networks with power law degree distribution with exponent $3<\gamma<4$, we present new value of critical exponent $\beta_{B}^{\mathcal{T}}=2/(3-\gamma)$. In fig.~\ref{fig:powerlaw_g33} in green line, only analytical results for the bond CAP with $\gamma=3.3$ and all $C= 3$  avoidable colors and equal frequencies are shown. Since large precision is needed to get criticality in this regime, we would need much bigger networks to get the probabilities that are small as calculated by analytics. However, uniqueness of critical exponents for scale free networks in typical regimes of power law exponents, makes CAP usfule for a deeper understanding of scale free networks in general.

In the following we will understand the origin of this result for the order parameter critical exponent $\beta_B^{\mathcal{T}}$. In order to proceed we need to define general coefficients $a_i$ $\forall i \in \mathbb{R}$, that will be used later in an extension of series expansion of our generating function, as 
\begin{equation}
    a_i = \frac{1 }{\left\langle k \right\rangle} \left\langle \frac{\Gamma (i+1-k)}{\Gamma (-k) \ \Gamma (i+1)} \right\rangle. 
\label{eq:coeff}
\end{equation}
The general definition of the coefficients $a_i$ is derived by using common relations for binomial coefficients\cite{graham1994concrete} and their connection with the gamma function\cite{garrappa2007some}. This coefficents are often used in percolation theory on scale free networks \cite{cohen2002percolation}. The sign of coefficients for $i \in \mathbb{N}$ alternates as $a_i = \left( - 1 \right)^{(i+1)} \left\langle \prod_{j=0}^{i} (k-j) \right\rangle / \left\langle k \right\rangle$. It is known~\cite{Newman_book} that the generalization of the statistical moments of power law distribution on $\left\langle k^m \right\rangle, \ m> 0, m \in \mathbb{R}$, diverges when $m \geq \gamma-1 $. Since this restriction has consequences on the percolation threshold, we shortly repeat them here concerning the threshold for the existence of the CAGC. First, the percolation threshold $\tilde{\phi}_{crit}$ is  well-defined\cite{cohen2000resilience,molloy1995critical,callaway2000network} when at least a first statistical moment exists, i.e. if and only if $\gamma > 2$. Second, a phase transition at a non-vanishing $\phi_{crit}$ exists if $\gamma >3$, i.e. when the second moment is finite. But it was also shown that the phase transition still occurs~\cite{albert2000error,callaway2000network,cohen2001breakdown} in a case of the extreme dilution ($\phi_{crit} = 0$) of scale-free networks with $\gamma \in \left\langle 2,3 \right]$. In the rest of the paper we will use a shorter notation for the critical parameter of scale free networks: $\phi_{crit} \equiv 1/a_1$.

We perform an expansion in small variables $v_{\mathcal{Q}}$ and $\delta$ (compare eq.~\ref{eq:vQ_2_uP} to eq.~\ref{eq:poi_gen_tay}), similar as we did before for Poisson graphs.  
The generating function of a power law degree distribution is not analytic around the argument of one, because there is always a higher derivative which diverges. Therefore, we use instead of the Taylor expansion an adjust version of the asymptotic form for the generating function of excess degree\cite{cohen2002percolation}: 
\begin{align}
    g_1(1-\epsilon) &= \frac{1}{\left\langle k \right\rangle} \sum\limits_{k = 0}^{\infty} k p_k \left[ 1- \epsilon \right]^{k-1} \nonumber \\
    &\approx  1 - a_1 \epsilon - a_2 \epsilon^2 -...- a_{\gamma-2 } \epsilon^{\gamma-2 },\label{eq:g1_series}\\   
    g_0(1-\epsilon) &\approx  1 - \left\langle k \right\rangle \epsilon + \frac{\left\langle k \right\rangle a_1}{2} \epsilon^2 + \frac{\left\langle k \right\rangle a_2}{3} \epsilon^3 +\dots\nonumber\\  &\qquad + \frac{\left\langle k \right\rangle a_{\gamma-2 }}{\gamma-1} \epsilon^{\gamma-1 },    
\label{eq:g0_series}
\end{align}
where coefficients $a_i,\ i>0, i \in \mathbb{R}$ are defined in eq.~\ref{eq:coeff}. Later we will replace $\epsilon$ with small variables $1-u_{\{q\}}$, as well as $1-u_{\{q,c\}}$ etc. Eq.~\ref{eq:g0_series} is derived by integrating $\left\langle k \right\rangle g_1(x)=\frac{\rm d}{{\rm d}x}g_0(x)$ from $x=1$ to $x=1-\epsilon$ and adjusting $g_0(1)=1$. Note that the term $a_{\gamma-2}$ implies that 
$\gamma > 2$, and therefore the first moment $\langle k\rangle$ is finite.  

Since the coefficients $a_i, \ i\in \mathbb{N}$ are of order $a_i \sim \langle k^{i-1} \rangle$, it follows that it must be $i \leq  \lfloor \gamma  -2\rfloor $, i.e. $\lfloor \gamma  -2\rfloor $ is the highest-order analytic term in the asymptotic expansion of the generating functions. If $\gamma \in \left\langle 2,3\right\rangle$, then only the first and the last term in eq.~\ref{eq:g1_series} exist (in eq.~\ref{eq:g0_series}, the first, second and last term exist). First, second and last term are expressed in eq.~\ref{eq:g1_series} if $\gamma \in \left\langle 3,4\right]$ (in eq.~\ref{eq:g0_series}, the first, second, third and last term are expressed). The case when $\gamma \geq 4$ fits with the usual mean field, since the first three terms in eq.~\ref{eq:g1_series} are sufficient to describe the behavior entirely. 

We plug expansion eq.~\ref{eq:g0_series} into eq.~\ref{eq:B_color}. With the small dual variables $v_{\{q\}}, v_{\{c\}}, v_{\{q,c\}}, \dots \ll 1$ as defined in eq.~\ref{eq:vQ_2_uP}, we find that many terms cancel out. For arbitrary avoidable subsets $\mathcal{T}$ we find
\begin{equation}
    B_{color}^{\mathcal{T}} \approx \langle k \rangle v_{\mathcal{T}}+{\rm HO}_{\gamma}.
\label{eq:B_color_v}
\end{equation}
If some of the nodes or edges are trusted, the linear term $\langle k \rangle v_{\mathcal{T}}$ dominates the expansion and therefore determines the critical exponent. The higher order terms ${\rm HO}_{\gamma}$ are only important, if all colors are avoided and therefore $v_{\mathcal{T}}=0$ holds.  We find 
\begin{equation}
    {\rm HO}_{\gamma} \sim \sum\limits_{\mathcal{Q} \subseteq \mathcal{T}} \left( -1 \right)^{\mid \mathcal{Q} \mid} (1-u_{\mathcal{Q}})^{\min(2,\gamma-1)}.
\label{eq:B_color_v_HO}
\end{equation}
Here, the small variables $(1-u_{\mathcal{Q}})$ can be replaced with the small variables $v_{\mathcal{P}}$, according to eq.~\ref{eq:vQ_2_uP}. 
So far we calculated the dominating lowest order terms of $B_{\rm color}^{\mathcal{T}}$ as a function of $v_{\mathcal{Q}}$. This expansion is valid close to the critical point. For identifying the critical exponent $\beta_B^{\mathcal{T}}$ (compare eq.~\ref{eq:critical_onset}), we still need to calculate how the small variables $v_{\mathcal{Q}}$ scale with $\delta=\phi-\tilde{\phi}_{crit}$. There are many interesting special cases of percolation transition for scale free graphs and we now list all of them.

\subsubsection*{One avoided color with $r_q<1$}

The first step is to solve the case for avoiding a single color $q$. The fraction of nodes or edges with this color has to be $r_q<1$. If all nodes or edges would be avoided, connectivity would be impossible. Avoiding only one color is identical with standard percolation~\cite{krause2017color}, but we need to go this first step before we can go to more avoided colors. The associated self-consistent equation, derived from eq.~\ref{eq:self_consistent} for $\mathcal{T} = \{ q\}$, is
\begin{equation}
\label{eq:uq_self}
    u_{\{q\}} = (1-\phi)+ \phi \left[r_q + (1-r_q) g_1(u_{\{q\}})\right],
\end{equation}
which is only a slightly modified self-consistent equation of regular percolation. 
After including the asymptotic expansion eq.~\ref{eq:g1_series} in the self-consistent relation eq.~\ref{eq:uq_self}, we get
\begin{align}
\label{eq:avoiding_q}
    ( \tilde{\phi}_{crit} +\delta )  &\left[ a_1  + a_2 v_{\{q\}} +...+ a_{ \gamma-2} v_{\{q\}}^{ \gamma-3 } \right] = \frac{1}{\left( 1-r_q\right)}.
\end{align}
The higher order terms in $\delta$ are further neglected, because only a linear term in $\delta$ is sufficient to get a behavior near criticality. The result of this equation for different $\gamma$-regimes follows directly 

\begin{align}
\label{eq:vq_crit}
    &v_{\{q\}} \approx  \nonumber \\
&\begin{dcases}
    \left[ (1-r_q) \ a_{\gamma-2} \delta \ \right]^{\frac{1}{3-\gamma}}, & \gamma \in \left\langle 2, 3 \right\rangle \\
    \left[ \frac{1}{\tilde{\phi}_{crit} a_{\gamma -2}} \left( \frac{1}{1-r_q} -a_1 ( \tilde{\phi}_{crit} +\delta ) \right) \right]^{\frac{1}{ \gamma-3 }}, & \gamma \in \left\langle 3, 4 \right] \\
    \frac{1}{ \tilde{\phi}_{crit} a_2 } \left( \frac{1}{1-r_q} -a_1 ( \tilde{\phi}_{crit} +\delta ) \right) , & \gamma \in \left[4, +\infty \right\rangle.
\end{dcases}
\end{align}
Note that the result for $\gamma \in \langle 2, 3 \rangle$ is determined with $\tilde{\phi}_{crit} = 0$. For other values of $\gamma$, the percolation threshold computed by eq.~\ref{eq:threshold} is such that  $(1-r_q)^{-1} - (\tilde{\phi}_{crit} + \delta) a_1 = -a_1 \delta$, only if color $q$ has maximal frequency $r_q$. If that is not the case, $v_{\{q\}}$ is critical at $\phi_{crit} / (1-r_q) < \tilde{\phi}_{crit}$. Solution of eq.~\ref{eq:vq_crit} for $\gamma = 3.3$ is presented in fig.~\ref{fig:powerlaw_g33}, noted with black line.

Returning to $B_{\rm color}^{\{ q\}}$, we use only the linear term of eq.~\ref{eq:B_color_v} with $v_{\mathcal{T}}=v_{\{q\}}$. 
With $B_{\rm color}^{\{ q\}} \sim \delta^{\beta_B^{\{ q\}}}$ we find the critical exponent for a colored scale-free network when only a single color is avoided 
($\mathcal{T} = \{q\}$) and this one avoided color covers only a part of all edges or nodes ($r_q<1$)
\begin{align}
\label{eq:beta_B}
    \beta_B^{\{ q\}} =
\begin{dcases}
    \frac{1}{\mid \gamma-3 \mid}, & \gamma \in \left\langle 2, 3 \right\rangle \cup \left\langle 3, 4 \right] \\
    1, & \gamma \in \left[4, +\infty \right\rangle.
\end{dcases}
\end{align}
This is a well-known result already reported for regular percolation~\cite{Newman_book}. In the fig.~\ref{fig:powerlaw} we demonstrate this behavior clearly. One has to note that the giant component does not exist for $\gamma > 3.478$ for usual scale free networks \cite{ChungLu} although it exists if minimal degree is controlled in the network. We report mean field exponents as we do not expect them to change in this pruned case.

\subsubsection*{Two avoided colors with $r_q+r_c=1$}

\begin{figure}[h!]
\includegraphics[scale=0.42]
{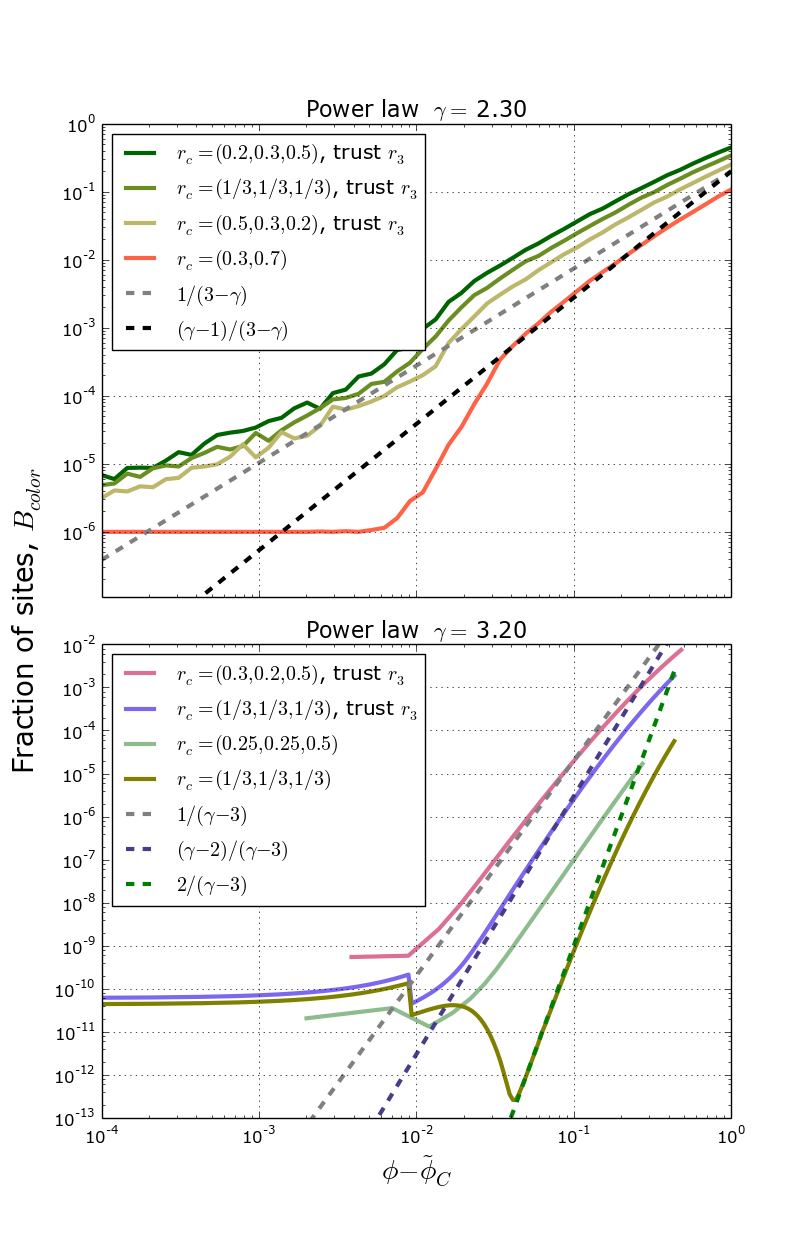}
\centering
\caption{ \textbf{Critical behavior of CAGC in edge colored scale-free network with trusted color and different color frequencies under bond percolation.} 
Power law degree distribution with exponents: $\gamma = 2.3$ (top) and $\gamma = 3.2$ (bottom). Other characteristics of simulations are the same as described in previous captions.
In top subfigure, \textit{green lines} represent cases of $C= 3$ colors when only the colors with $r_1, r_2$ are avoided. Shades of \textit{green} from darker to lighter show three possible relations between colors. \textit{Orange line} are simulations of $C= 2$ avoiding colors, but with different frequencies $(r_1, r_2) = (0.3, 0.7)$. \textit{The dashed black line} is a power law with exponent $(\gamma-1)/(3-\gamma)$, and \textit{the dashed gray line} with $1/(3-\gamma)$. 
In bottom subfigure, $4$ different scenarios of trustiness and relations between color frequencies (compare with table~\ref{tab:all_cr_exp}) are compared with simulations of networks with $C=3$. \textit{Purple lines} are used when one color is trusted (color with frequency $r_3$), while case when all colors are avoidable (untrusted) are shown with \textit{green lines}. Equal frequencies of avoidable colors is noted in darker shades. Three different critical exponents are found -- $1/(3-\gamma)$, noted again with \textit{dashed gray line}, $(\gamma-2)/(\gamma-3)$ noted with \textit{dashed purple line} and $2/(\gamma-3)$ with \textit{dashed green line}. 
} 
\label{fig:powerlaw_hetero_trust}
\end{figure}

When a number of colors are avoided, the presented algorithm is repeated, but a significant difference from regular percolation sets in. Let us start with the simplest case, where only two colors are present in the network and these two colors are untrusted ($\mathcal{T}=\{q,c\}$) and have the same shares $r_q=r_c$. We know that $v_{\mathcal{T}}=0$ and $v_{\{q\}}=v_{\{c\}}$, therefore we find with eq.~\ref{eq:B_color_v_HO} for the leading order 
\begin{align}
B_{\rm color}^{\mathcal{T}} &\sim -2 (v_{\{q\}})^{\min(2,\gamma-1)} + (2 v_{\{q\}})^{\min(2,\gamma-1)}\nonumber\\
&\sim (v_{\{q\}})^{\min(2,\gamma-1)}.
\end{align}
Together with eq.~\ref{eq:vq_crit} we find the critical exponent for $C=2$ colors which are both avoided and have identical shares $r_q=r_c$:
\begin{align}
    \beta_B^{\{q,c\}} =
\begin{dcases}
    \frac{\gamma-1}{3 - \gamma}, & \gamma \in \left\langle 2, 3 \right\rangle  \\
    \frac{2}{\gamma -3},  & \gamma \in \left\langle 3, 4 \right]           \\
    2, & \gamma \in \left[ 4, +\infty \right\rangle.
\end{dcases}
\label{eq:beta_b_untrused_homo_two}
\end{align}
This result is confirmed in fig.~\ref{fig:powerlaw} with $\gamma=2.3$, see the red line and red triangles on the right of the figure.  With green line on fig.~\ref{fig:powerlaw_g33} we show result for $\gamma=3.3$.

When there is a difference in the color frequencies ($r_c > r_q$), the higher order terms of eq.~\ref{eq:B_color_v_HO} contain the terms of mixed order, like in the case of Poisson distribution in eq.~\ref{eq:poi_expansion}. Terms of the type $v_{\{ q\}}v_{\{ c\}}$ in that case, are of lowest order in $\delta$. In that way criticality of $B_{\rm color}^{\mathcal{T}}$ is determinated by different thresholds of probabilities $v_{\{ q\}}$ and $v_{\{ c\}}$ and only for $\gamma>3$ the threshold is larger than zero and the results are affected by difference in color frequencies. The leading order for heterogeneous coloring is thus 
\begin{align}
    \beta_B^{\{q,c\}} =
\begin{dcases}
    \frac{\gamma-1}{3 - \gamma}, & \gamma \in \left\langle 2, 3 \right\rangle  \\
    \frac{1}{\gamma -3},  & \gamma \in \left\langle 3, 4 \right]           \\
    1, & \gamma \in \left[ 4, +\infty \right\rangle.
\end{dcases}
\label{eq:beta_b_untrusted_hetero_two}
\end{align}
The behavior of CAC in the case of unequal color distributions is presented in fig.~\ref{fig:powerlaw_hetero_trust}.

\subsubsection*{Two avoided colors with $r_q + r_c < 1$} 

If the third color, which can be trusted, exists, a next step of iteration in solving self-consistent eq.~\ref{eq:self_consistent} is needed. More precisely, the self-consistent equation for $u_{\{q,c\}}$ is derived from eq.~\ref{eq:self_consistent} for $\mathcal{T} = \{q,c\}$, 
\begin{align}
\label{eq:u_cq_self}
    u_{\{q,c\}} = (1-\phi)+ &\phi \left[r_c g_1(u_{\{q\}}) + r_q g_1(u_{\{c\}})  \right. \nonumber \\ 
    &\left. + (1-r_q-r_c) g_1(u_{\{q,c\}})\right].
\end{align}
It depends on the generating functions of excess degree $g_1(u_{\{q\}})$ and $g_1(u_{\{c\}})$, as they were used already in the previous iteration step for one avoided color. Additionally, a new generating function of excess degree $g_1(u_{\{q,c\}})$ is included when $r_q+r_c<1$. The self-consistent equation for $g_1(u_{\{q\}}), g_1(u_{\{c\}})$ in eq.~\ref{eq:uq_self} are modified using eq.~\ref{eq:uQ_2_vP} and eq.~\ref{eq:phi_c_delta}, so a relation
\begin{equation}
    g_1(u_{\{q\}}) = 1 - \frac{v_{\{q\}}}{\left( \tilde{\phi}_{crit} + \delta \right) \left(  1-r_q \right)},
\label{eq:uq_self_2_v}
\end{equation} 
is used in the further calculation. 

In the same way, the self-consistent equation for avoiding two colors in eq.~\ref{eq:u_cq_self}, which is relevant for the case $r_q+r_c<1$ is rewritten in terms of variables $v$ as
\begin{align}
    g_1(u_{\{q, c\}}) &= 1 - \frac{v_{\{q\}}}{\left( \tilde{\phi}_{crit} + \delta \right) \left(  1-r_q \right)} - \frac{v_{\{c\}}}{\left( \tilde{\phi}_{crit} + \delta \right) \left(  1-r_c \right)} \nonumber \\
    &+ \frac{v_{\{q,c\}}}{\left( \tilde{\phi}_{crit} + \delta \right) \left(  1-r_q -r_c \right)}.
\label{eq:uqc_self_2_v}
\end{align} 
Note that we use the asymptotic form of excess generating function for $\epsilon = u_{\{q,c \}} = v_{\{c\}}+v_{\{q\}}-v_{\{q,c\}}$ in eq.~\ref{eq:g1_series}. Combined with eq.~\ref{eq:uq_self_2_v}, we get the asymptotic form of self-consistent equation for $v_{\{q,c\}}$, which is in analogy with eq.~\ref{eq:avoiding_q} but now for avoiding two colors

\begin{align}
    &\frac{v_{\{q,c\}}}{\left( \tilde{\phi}_{crit} + \delta \right) \left(  1-r_q -r_c \right)} \approx - 2 a_2 v_{\{q\}} v_{\{c\}} \nonumber \\
    & +\left[ a_1 + 2 a_2 \left( v_{\{q\}} + v_{\{c\}} \right) \right]  v_{\{q,c\}} - \dots \nonumber \\
    &- a_{\gamma-2} \left[ \left( v_{\{q\}} + v_{\{c\}} - v_{\{q,c\}}\right)^{\gamma-2} - v_{\{q\}}^{\gamma-2} - v_{\{c\}}^{\gamma-2} \right].
\label{eq:avoiding_qc}
\end{align}

To describe the criticality of $v_{\{q,c\}}$, once again only linear order terms of $\delta$ are used. In the second step of iteration, a new cross-product terms of $v_{\{q \}}$, $v_{\{c \}}$ and $v_{\{q,c \}}$ appear in calculation and their interrelation is important for understanding of criticality. Using eq.~\ref{eq:vq_crit} one notices that condition $r_c > r_q$, implies that when $v_{\{q \}}$ is critical, $v_{\{c \}}$ is not, because they have different thresholds. Therefore $v_{\{q,c \}}$, which is the probability of site being connected to the giant component when both colors are simultaneously avoided, is critical when both $v_{\{q \}}$ and $v_{\{c \}}$ are non-zero and at least one of them is critical. For our case, $r_c > r_q$ means that critical threshold for $v_{\{q,c \}}$ is equal to the one of $v_{\{c \}}$. Furthermore, that is the reason why and when some of the cross-product terms are not contributing to criticality of $v_{\{q,c \}}$. 

In eq.~\ref{eq:avoiding_qc}, only $v_{\{q\}} v_{\{c\}}$ product has effect on criticality of $v_{\{q,c \}}$ and this happens only in the mean field regime $\gamma \geq 4$. Other mixed linear terms are corrections of higher order in criticality. This is in accordance with our results previously shown for Poisson graphs. For other $\gamma$ exponents, the term with coefficient $a_{\gamma -2}$ is necessary to describe the criticality of networks with long-tail degree distributions, just as in regular percolation. For $\gamma \in \left\langle 2, 3 \right\rangle$ only the $a_{\gamma -2}$ term remains, just as in the case of one color avoidance eq.~\ref{eq:avoiding_q}. For $\gamma \in \left\langle 3, 4 \right\rangle$, the contribution of the term with coefficient $a_1$ is included to the $a_{\gamma -2}$ term. When dealing with the $a_{\gamma -2}$ term, we use approximation $v_{\{c\}} - v_{\{q,c\}} \ll v_{\{q\}}$ together with $r_c > r_q$, such that 
\begin{align}
\left( v_{\{q\}} \right. +& \left. v_{\{c\}} - v_{\{q,c\}}\right)^{\gamma-2} \approx \nonumber \\
& v_{\{q\}}^{\gamma-2} \left( 1 + (\gamma -2) (v_{\{c\}} - v_{\{q,c\}} / v_{\{q\}} ) \right).
\label{eq:approx}
\end{align} 
This approximation is based on the property that  $v_{\{q,c\}} << v_{\{q\}}$, as we show in fig.~\ref{fig:powerlaw_g33} with black and purple lines.
 
After the approximation in eq.~\ref{eq:approx}, only $v_{\{q\}}^{\gamma-3} v_{\{c\}}$ product remains as the mixed term with coefficient $a_{\gamma-2}$ of the lowest order in $\delta$. From eq.~\ref{eq:avoiding_qc} it finally follows for $r_c > r_q$ 

\onecolumngrid

\begin{align}
    v_{\{q,c\}} = 
\begin{dcases}
    \left[ (1-r_q-r_c) \ a_{\gamma-2} \delta \ \right]^{\frac{1}{3-\gamma}}, &\qquad \gamma \in \left\langle 2, 3 \right\rangle \\
    \frac{ - (\gamma-2) \left( \frac{1}{1-r_q} - a_1 \tilde{\phi}_{crit} \right)}{ \frac{1}{1-r_q- r_c} -  a_1 \tilde{\phi}_{crit} - \frac{\gamma-2}{a_{\gamma-2}} \left( \frac{1}{1-r_q} - a_1 \tilde{\phi}_{crit} \right)} \left[ \frac{-a_1}{\tilde{\phi}_{crit} a_{\gamma-2}} \delta \right]^{\frac{1}{ \gamma-3}},& \qquad \gamma \in \left\langle 3, 4 \right\rangle \\
    \frac{ -2 a_1 \left[ a_1 \delta^2 + \left( 2 \tilde{\phi}_{crit} a_1 - \frac{1}{1-r_q} - \frac{1}{1-r_c} \right) \delta \right] }{\tilde{\phi}_{crit} a_2 \ \left(\frac{1}{1-r_q-r_c} - \tilde{\phi}_{crit} a_1 \right) } ,& \qquad \gamma \in \left[4, \infty \right\rangle.
\end{dcases}
\label{eq:vqc_crit}
\end{align}
\twocolumngrid


Note that the crucial mixed term remains for $r_q = r_c$ in all $\gamma$-regimes. But only for $\gamma \in \left\langle 3,4 \right\rangle$ that changes the critical exponent for $v_{\{q,c\}}$, as shown in fig.~\ref{fig:powerlaw_g33} with purple line. For that regime of $\gamma$, the mixed term is in form of $v_{\{q\}}^{\gamma-2}$ and it follows that 
\begin{align}
v_{\{q,c\}} = \frac{- (\gamma-2)}{\frac{1}{1-2r_q} - a_1 \tilde{\phi}_{crit}} \left( \frac{1}{\tilde{\phi}_{crit} a_{\gamma-2}} \right)^{\frac{1}{ \gamma-3}} \left( -a_1 \delta \right)^{\frac{\gamma-2}{\gamma-3}}.
\label{eq:vqc_vrit}
\end{align}

The difference between criticality of probability $v_{\{q,c\}}$ in \ref{eq:vqc_crit} and \ref{eq:vqc_vrit} is clearly related to the ratio of avoided color frequencies. Although this behavior still depends only on power law exponent $\gamma$, it is also marked as a consequence of the existence of color on network.

The result for $\gamma \in \langle 2, 3 \rangle$ is again determined with $\tilde{\phi}_{crit} = 0$. In other regimes, the percolation threshold is given with the condition that zero-order term vanishes. 
In the regular mean field regime $\gamma \geq 4$, the before mentioned condition generates the quadratic equation 
\begin{equation}
    a_1^2 \tilde{\phi}_{crit}^2 - a_1 \tilde{\phi}_{crit} \left( \frac{1}{1-r_q} + \frac{1}{1-r_c}\right) + \frac{1}{(1-r_q)(1-r_c)} = 0
\end{equation}
with results $\tilde{\phi}_{crit} = \left( (a_1 (1-r_c))^{-1}, (a_1 (1-r_q))^{-1} \right)$. If we select the first one ($r_c>r_q$), the result from eq.~\ref{eq:vqc_crit} for $\gamma \in \left[4, \infty \right\rangle$ becomes 
\begin{align}
    v_{\{q,c\}} = \ \frac{- 2 a_1^2}{a_2} & \frac{(1-r_c)^2 (1-r_q-r_c)}{r_c} \left[ a_1 \delta ^2  \vphantom{\frac{1}{(1+1)}} \right. \nonumber\\
    & \left. + \ \frac{r_c-r_q}{(1-r_q)(1-r_c)} \delta \right]
\label{eq:vqc_critic_MFT}
\end{align}
Note that the linear term in $\delta$ gives physical solution when $r_c > r_q$. If $r_q = r_c$, then only the quadratic term remains. In this regime, critical exponent is equal to the number of dominant colors as found in previous works for ER networks~\cite{krause2016hidden,krause2017color}.

In the presence of some trusted colors,
the higher order terms in eq.~\ref{eq:B_color_v} can be neglected and the relative size of the CAGC scales linear with $v_{\{q,c\}}$. Therefore we find for two colors with identical shares $r_q=r_c<1/2$

\begin{align}
    \beta_B^{\{q,c\}} =
\begin{dcases}
    \frac{1}{3 - \gamma}, & \gamma \in \left\langle 2, 3 \right\rangle \\
     \frac{\gamma-2}{\gamma-3}, & \left\langle 3, 4 \right] \\
    2, & \gamma \in \left[4, +\infty \right\rangle
\end{dcases}
\label{eq:beta_b_result_trusted_homo_two}
\end{align}
and for different color frequencies $r_q \neq r_c, \ r_q, r_c<1/2$
\begin{align}
    \beta_B^{\{q,c\}} =
\begin{dcases}
    \frac{1}{\mid \gamma-3 \mid}, & \gamma \in \left\langle 2, 3 \right\rangle \cup \left\langle 3, 4 \right\rangle \\
    2, & \gamma \in \left[4, +\infty \right\rangle.
\end{dcases}
\label{eq:beta_b_result_trusted_hetero_two}
\end{align}

Note that if trusted colors exist in hubs-like network (power law with $\gamma \in \left\langle 2, 3 \right\rangle \cup \left\langle 3, 4 \right\rangle$), CAGC is in fact dominated by the giant component of the trusted part of the network. That is the reason why the critical exponent of CAGC is equal to the one for the ordinary giant component. The behavior of CAC in the cases with trusted color and different color distributions are presented in fig.~\ref{fig:powerlaw_hetero_trust}.

\subsubsection*{General results}

The critical exponent of the order parameter for networks with two avoided colors $\mathcal{T}=\{q,c\}$ can be generalized. We induce it for arbitrary color-avoiding sets $\mathcal{T}$, and in the case of bond percolation all possible situations are summerized in table~\ref{tab:all_cr_exp}. This results are confirmed in fig.~\ref{fig:powerlaw} with $\gamma=2.3$ and $T=|\mathcal{T}|=10$ avoided colors, fig.~\ref{fig:powerlaw_g33} with $\gamma=3.3$ and $T=3$ and fig.~\ref{fig:powerlaw_hetero_trust}.

\begin{table}[htbp]
    \centering
    \begin{tabular}{c | c | c}
    \textbf{$\beta_B^{ \mathcal{T}}$} & \textbf{Trusted} & \textbf{Untrusted} \\
    \hline
    \textbf{Equal color frequencies} &  &  \\
    $\gamma \in \left\langle 2, 3 \right\rangle$ & $1/(3 - \gamma)$ & $(\gamma-1)/(3 - \gamma)$ \\
    $\gamma \in \left\langle 3, 4 \right\rangle$ & $(\gamma-2)/(\gamma -3)$ & $2/(\gamma -3)$ \\
    $\gamma \in \left[4, +\infty \right\rangle$ & $C$ & $|\mathcal{T}| < C$ \\
    \hline    
    \textbf{Different color frequencies} &  &  \\
    $\gamma \in \left\langle 2, 3 \right\rangle$ & $1/(3 - \gamma)$ & $(\gamma - 1)/(3 - \gamma)$ \\
    $\gamma \in \left\langle 3, 4 \right\rangle$ & $1/(\gamma -3)$ & $1/(\gamma -3)$ \\
    $\gamma \in \left[4, +\infty \right\rangle$ & $|\mathcal{Q}|$ & $|\mathcal{Q}|$ \\
    \end{tabular}    
\caption{\textbf{Critical exponent for the bond color-avoiding percolation on scale free networks.} Size of the giant color-avoiding component (order parameter) behaves like $B_{color}^{ \mathcal{T}} \sim \left( \phi - \tilde{\phi}_{crit} \right)^{\beta_B^{ \mathcal{T}}}$ near criticality. Frequencies of colors $r_i$ and existence of trusted colors determine the criticality, but differently for each regime of power law exponent $\gamma$. While the criticality is dominated by existence of colors for usual mean field regime $\gamma \in \left[4, +\infty \right\rangle$, in other regimes it is driven by hub-like network topology. The avoidance of color makes distinction even between networks with hub-like topology, which is not the case in regular percolation. With $|.|$ we note number of colors in set of avoiding colors $\mathcal{T}$, or in set of avoiding colors with the highest frequency in $\mathcal{Q} \subset \mathcal{T} : r_q = \mathrm{max}_{t \in \mathcal{T}}r_t,\ \forall q \in \mathcal{Q}$.}
\label{tab:all_cr_exp}
\end{table}

\begin{figure}[h]
\includegraphics[scale=0.46]{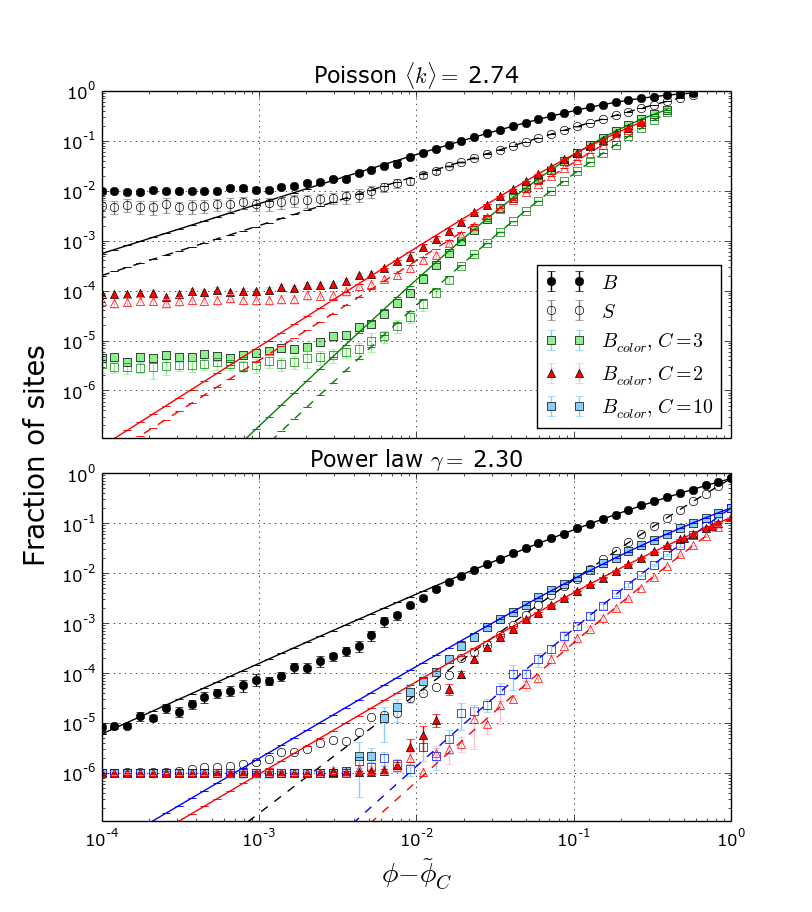}
\centering
\caption{ \textbf{Bond and site percolation on edge colored network.} Results for the sizes of giant component and CAGC obtained by simulations are shown in symbols and the one from our theory in lines. \textit{Full symbols} and \textit{solid lines} are for bond percolation (repeated from fig.~\ref{fig:poisson} and fig.~\ref{fig:powerlaw} in the same choice of colors), while \textit{empty symbols} and \textit{dashed lines} present results for site percolation also with the same associated colors. Erd\"os-R\'enyi (up) and scale-free (down) network have equal average degree. For CAGC, no matter how many different colors are in the network, we see that, while $\beta_S = \beta_B$ stays in Poisson case, in the power law with $\gamma = 2.3$ emerges the breaking of the site-bond percolation universality, since $\beta_S = \beta_B+1$. The breaking of site-bond percolation universality was already noted but for the giant component\cite{radicchi2015breaking}, here shown in \textit{black}.}
\label{fig:v_e_percolation}
\end{figure}

As a main result in this paper, we show that the critical exponent of scale free colored networks is dependent on the number of possibly vulnerable colors (avoidable colors) only in the usual mean-field regime, while in the unusual mean-field regime stays dominantly affected as a topological consequence of the networks long-tail degree distribution. 
Interestingly the critical exponent depends on the presence of colors for $\gamma \in \left\langle 2,4 \right\rangle$, although it depends only on the exponent of power law degree distribution. For $\gamma \in \left\langle 2, 3 \right\rangle$ we get that CAP critical exponent is not sensitive to different frequencies of colors (heterogeneous coloring). On the contrary, it does depend on the existence of a trusted color. Trusted color decreases the CAP critical exponent on scale-free networks such that it is equal to the one of standard percolation. 
For $\gamma \in \left\langle 3, 4 \right\rangle$ we see that CAP critical exponent grows with the level of degeneracy of highest color frequency. Dependence on color frequencies is still driven by topology, but differently then for mean field or other topological regimes. In this regime, the dependence on trusted color is noticeable only if color frequencies are equal. Further, critical exponent is still different for homogeneous and heterogeneous coloring even when the trusted color exist.

Color dependent and color independent critical behaviors in colored networks are clearly demonstrated in comparison of right panels in fig.~\ref{fig:poisson} and fig.~\ref{fig:powerlaw}. Our analytically calculated critical exponents are in agreement with numerical results. Also note that outside of critical regime, networks with $\left\langle 2, 3 \right\rangle \cup \left\langle 3, 4 \right\rangle$ are still color dependent. Note also that assumption of the same order of $v_{\{q\}}, v_{\{c\}}, v_{\{q,c\}}$ is vindicated with excellent match between simulations and theory. The intuition for this behavior is that for inhomogeneous coloring they are of the same order while in the case of homogeneous coloring ($r_q=r_c$), the joint probability term is $\sim\delta^2$. 

In the case in which the color distribution varies for different colors (heterogeneous coloring), from eq.~\ref{eq:vqc_critic_MFT} it is clear that the dominant color defines critical behavior, while critical exponent stays equal to single-color-avoiding case in eq.~\ref{eq:beta_B}. If there are few colors which share their dominance, the critical exponent is equal to the number of degeneration of color-vector components. Such behavior was also observed previously~\cite{krause2017color}. To the contrary, when coloring is homogeneous ($r_q = r_c, \ \forall q,c \in \mathcal{T}$), mixed term of type $\prod_{q \in \mathcal{T}} v_{\{q\}}$ raises critical exponent to be equal to the number of colors that are avoidable, but only in the usual mean filed regime. This is already known behavior~\cite{krause2017color}, but result here emphasizes that in this respect the edge coloring behaves the same as vertex coloring. This is true only in mean field $\gamma \leq 4$. Heterogeneous coloring will change critical exponent for power law $\gamma \in \left\langle 3, \infty \right\rangle$, where threshold is non-null.

An important advantage of the use of $\phi$ as the percolation parameter, instead of $\langle k \rangle$ which was used in previous papers on color avoiding percolation, is that it is not depending on the choice of the degree distribution. The choice of $\langle k \rangle$ for the percolation seems very natural for the Poisson distribution since its parameter is exactly the average degree. Instead, in power law distribution, the relation between average degree and its parameter $\gamma$ is such that $\langle k \rangle$ is barely changing when $\gamma \gtrsim 4$. That is the reason why parameters in correspondence with the degree distribution are not sufficient to explain appropriately the critical behavior. However, the percolation threshold is not affected by this situation, since for both choices it remains dependent on the degree distribution. On the other hand, limitations of our simulations of synthetic networks appear at $\gamma \gtrsim 3$ for $N  = 10^6$, where the CAGCs can not be simulated. We suggest that the reason is similar as for the limit for giant component~\cite{radicchi2014underestimating} but in this case it is even more constraining.

Except for the averaging with the different network configurations where each of them is colored by random, we also tried another possible averaging. We studied the numerical results averaged over the same network configuration but with different color realizations. In the first type of averaging the connectivity disorder is annealed and in second quenched, while the color stays as annealed property of edges in both cases. This results are not shown here, since there is no difference between those two averages, but we highlight that our theory covers both cases. However, we find interesting that the CAGC considered here is already in a replica symmetric phase~\cite{zdeborova2007phase}. Better understanding of this result is a subject of future work.

\section{Color-avoiding site percolation}

\begin{figure}[h]
\includegraphics[scale=0.4]{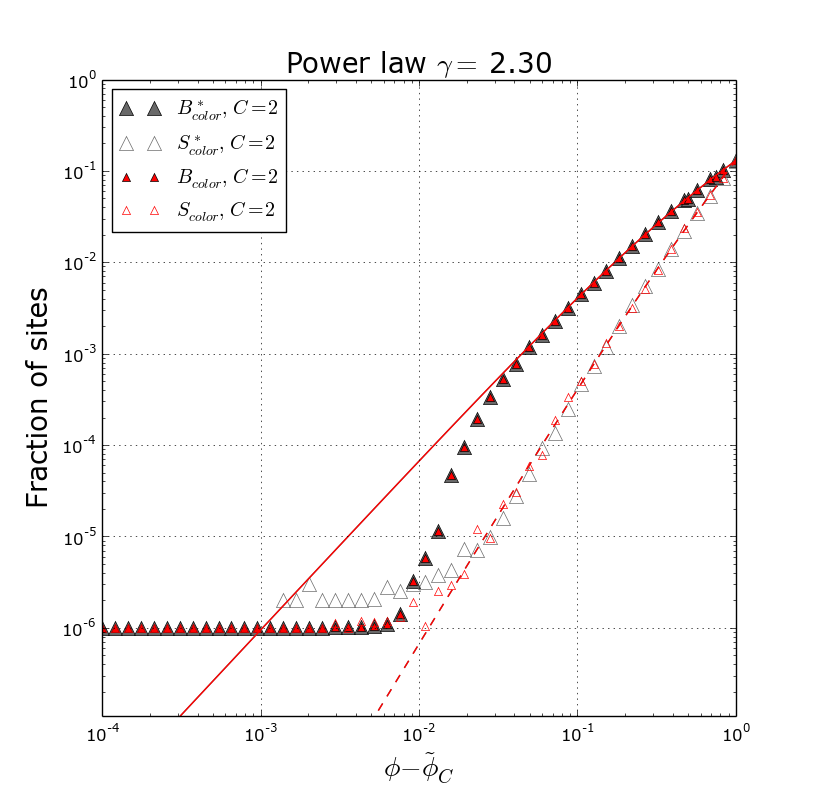}
\centering
\caption{ \textbf{Size of color avoiding connected component is equal for edge and vertex coloring under both site and bond percolation.} Numerical and analytical results are shown only for scale-free network with $\gamma = 2.3$ and $C= 2$ avoiding colors. Same as in supplements of fig.~\ref{fig:v_e_percolation}, \textit{full triangles} and \textit{solid line} are for bond percolation, while \textit{empty triangles} and \textit{dashed line} present results for site percolation. While results for edge coloring are shown in \textit{red}, results for size of CAGC in  the vertex coloring are noted in \textit{gray}. Other details are identical as in supplements of fig.~\ref{fig:powerlaw} and fig.~\ref{fig:v_e_percolation}. We see that critical exponents does not depend weather the color is edge or vertex property. One can better understand that by noting that the same situation happens while avoiding the color of edge from randomly chosen vertex or the color of its first neighbor.}
\label{fig:v_e_color}
\end{figure}

We find that the color-avoiding percolation strengths for bond percolation $B_{color}^{\mathcal{T}}$ and site percolation $S_{color}^{\mathcal{T}}$ in the edge coloring case are related by
\begin{equation}
    S_{color}^{\mathcal{T}} = \phi B_{color}^{\mathcal{T}},
\label{eq:SB_color_relation}
\end{equation} 
which is the same linear proportionality as in a regular percolation~\cite{radicchi2015breaking}. For Poisson graphs $\tilde{\phi}_{crit}>0$, and therefore the site and bond percolation have exactly the same exponents. This behavior is clearly presented in the upper part of fig.~\ref{fig:v_e_percolation}.

The breaking of the site-bond percolation universality follows for power law distributed networks only with $\gamma \in \left\langle 2, 3 \right\rangle$, because $\tilde{\phi}_{crit} = 0$. The associated critical exponent $\beta_S^{\mathcal{T}}$ of site color-avoiding percolation for the edge coloring is
\begin{align}
    \beta_S^{ \mathcal{T}}=
\begin{dcases}
    \frac{\gamma-1}{3-\gamma}+1, & \gamma \in \left\langle 2, 3 \right\rangle \\
    \frac{2}{\gamma-3}, & \gamma \in \left\langle 3, 4 \right\rangle \\
    \mid \mathcal{T} \mid, & \gamma \in \left[4, +\infty \right\rangle.
\end{dcases}
\label{eq:beta*_S*_result}
\end{align}
Presented are equations for all avoidable colors and equal frequencies, but the breaking of site and bond universality will be the case in any instance mentioned in table~\ref{tab:all_cr_exp} for regime $\gamma \in \left\langle 2,3 \right\rangle$. 

For vertex coloring, noted with $*$, we find that $\beta_S^{* \mathcal{T}} = \beta_S^{ \mathcal{T}}$. In the fig.~\ref{fig:v_e_color} we demonstrate that behavior of colored vertices and edges is exactly the same.

\section{Conclusion}
        
In conclusion, we have shown that in the regular mean field limit networks sense colors in color avoiding percolation and exponent of the percolation is always equal to the number of dominant colors. Topological constraints of scale-free networks are stronger then the color avoiding property and the critical behavior outside of the mean field regime depends on the color existence although computed critical exponents depend only on the exponent $\gamma$ of power law degree distribution. We have also shown that the site and bond percolation differ as was previously found. 
We believe that the study of color avoiding percolation is fundamentally important for studies of critical behavior in networks as they naturally interpolate between different interesting critical regimes.

\section{Acknowledgement}
        
We want to express ou gratitude to Anna Maria Loguericio for all the help during research and to Antonio Scala and Claudio Conti for all the fruitful discussions we had. VZ and GC acknowledge the Israel-Italian collaborative project NECST. AK and VZ acknowledge support by the H2020 CSA Twinning project No. 692194, “RBI-T-WINNING”. VZ acknowledge partial support form Croatian centers of excellence Quantix-lie and Center of Excellence for Data Science and Cooperative Systems, and European Commission FET-Proactive project MULTIPLEX (Grant No. 317532). AK acknowledges support of Croatian Science Foundation.

\bibliography{edge_coloring}

\end{document}